# Reply to Comment on "Tip induced unconventional superconductivity on Weyl semimetal TaAs"

He Wang[1,2,†], Huichao Wang[1,2,†], Yuqin Chen[1,2,†], Jiawei Luo[1,2], Zhujun Yuan[1,2], Jun Liu[3], Yong Wang[3], Shuang Jia[1,2], Xiong-Jun Liu[1,2*], Jian Wei[1,2*], and Jian Wang[1,2*]

[1]International Center for Quantum Materials, School of Physics, Peking University, Beijing 100871, China

[2]Collaborative Innovation Center of Quantum Matter, Beijing, China

[3]Center of Electron Microscopy, State Key Laboratory of Silicon Materials, Department of Materials Science and Engineering, Zhejiang University, Hangzhou 310027, China

[†]These authors contributed equally to this work.

*e-mail: jianwangphysics@pku.edu.cn; weijian6791@pku.edu.cn; xiongjunliu@pku.edu.cn

Recently, we presented a paper about the tip-induced superconducting phase on TaAs single crystal[1] (arXiv:1607.00513). A conductance plateau with double conductance peaks at finite bias, sharp double dips and a zero bias conductance peak were observed in the point contact spectra (PCS). These three features in one PCS suggest the possibility of p-wave like superconductivity and Majorana zero modes in the surface. Soon after our paper posted on arXiv, Gayen *el al.* (arXiv:1607.01405)[2] questioned the underlying mechanism of our observations based on their experimental measurements in conventional superconductors(Pb and Nb) and conclusively indicated that the superconductivity observed in our TaAs crystal is conventional. We disagree with their quick conclusion without any specific study on the TaAs materials. We will explain in the following that for the clean (Z=0) point contact (PC) in ballistic limit, the superconducting transition feature in temperature dependence of PC resistance is expected. And the critical current effect claimed by Gayen *el al.* cannot explain the observed PCS in our paper.

In the following we make a one by one response to all the issues raised by Gayen *et al.*, and point out the apparent misinterpretations in their comment.

***"It is well known that it is possible to obtain a tip-induced superconducting (we named it "TISC") phase in topologically non-trivial materials under mesoscopic point contacts. This was first shown by Aggarwal et al. in the preprint arXiv:1410.2072 (October, 2014), a modified version of which was eventually published in Nature Materials[6] and following that work other***

*groups reproduced the same effect (arXiv:1501.00418 (January, 2015), Nat. Mat. (2016))"*

The above claim is not appropriate. We two groups independently studied $Cd_3As_2$ by using hard point contact spectroscopy and reported tip induced superconductivity on $Cd_3As_2$ in Nature Materials simultaneously[3,4]. One major difference between our studies is the sample quality. Single crystalline $Cd_3As_2$ samples were used in our work [3], while polycrystalline samples were used in their work[4]. The single crystal $Cd_3As_2$ is a 3D topological Dirac semimetal with linear dispersion for electronic structure and exotic transport properties, which have been demonstrated and confirmed by previous ARPES[5-7], STM[8] and transport studies from various groups[9,10] including our group[11]. $Cd_3As_2$ single crystal is an ideal platform to realize topological superconductivity since 3D topological Dirac semimetal locates on the phase boundary of various topological phases[12]. However, so far there is no experimental evidence that a polycrystalline $Cd_3As_2$, as considered in work by Aggarwal *et al.*[4] can be topologically nontrivial.

*"the point contact resistance is highly temperature dependent and clearly shows the superconducting transition indicating that the bulk resistivity contributes significantly in the point contact resistance"*

Gayen *et al.* claimed that if there is a superconducting transition in the temperature dependence of PC resistance, it would indicate the PC is in the thermal limit. This is apparently at odds with the common knowledge of superconducting PC in the ballistic regime. If a PC made between two normal metal electrodes is in the ballistic limit, its resistance would be independent of the temperature. But as can be inferred from the well-known BTK theory[13], when a PC made between a superconductor and a normal metal is in the ballistic limit, its resistance would drop to half of its normal state value due to Andreev reflection when temperature decreases below the transition temperature of the superconductor. This resistance drop is clearly demonstrated in the theoretical curves (Fig. 1a), which are consistent with the experimental curves (Fig.1b). Similar figures can be found in Fig.6a in *Ref.* 14 and Fig. 4a in *Ref.* 15 too.

In Fig. 2a of our work (arXiv: 1607.00513), the PC resistance decreases from 18.8 to 17.4 Ω in the zero field cooling process. The PC resistance at low temperature (17.4 Ω) is larger than half of its normal state value (9.2 Ω), which we believe is due to a finite barrier (Z>0) at the interface of PC that suppresses the Andreev reflection. Our results can be reasonably interpreted by the BTK model with a finite barrier, so their claim that temperature dependent PC resistance exclusively indicates the contribution from bulk resistivity is incorrect.

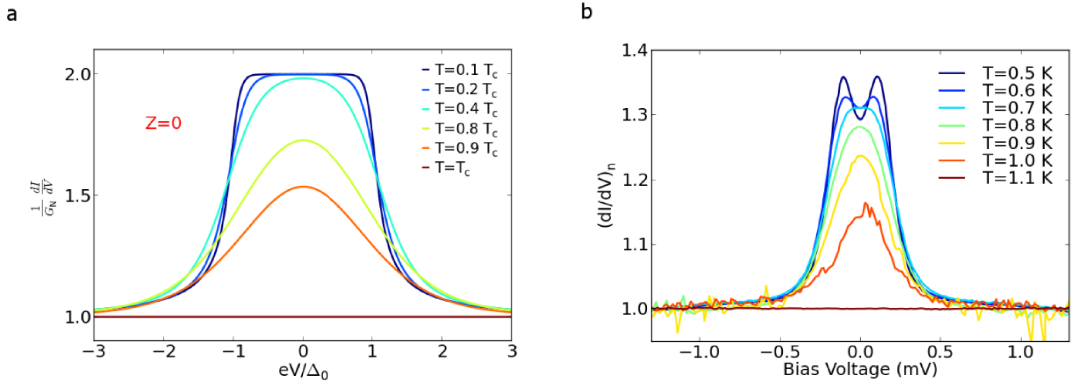

Figure 1. The zero bias conductance has clear temperature dependence for the PC in ballistic limit. a. The normalized conductance vs. normalized bias voltage curves at different temperatures with the barrier parameter Z=0 by BTK theory. The resistance at zero bias would drop rapidly to half of its normal state value when temperature decreases below $T_c$. b. Temperature dependence of the experimental normalized conductance curves for a PC made between a gold tip and an $Au_2Pb$ single crystal, which are comparable to the curves in Fig.1a. The same experimental data were presented in Fig. 3c of *Ref.*16.

*"First, they have not explained why they believe that the assumption $R_{PC} = R_{Sh} = 18.8\ \Omega$ is valid."*

We clarify that we never made such an assumption in our paper.

*"the authors started their discussion saying that the point contact is ballistic only when the contact size is less than the normal state mean free path of the sample. However, they have not provided such a comparison"*.

We didn't provide such a comparison because the Sharvin formula is not suitable to estimate the radius of a PC with a finite barrier (see detailed discussion in supplementary information of our arXiv paper[1] or in Appendix C in *Ref.* 17). So we proposed a new way to analyze the contribution of Maxwell resistance to the total resistance, which suggests the two PCs in our paper are close to the ballistic limit (supplementary information of our arXiv paper[1]).

*"the authors have not discussed why they believed that the magnetic field dependence would originate only from the sample, Rationally, the point contact itself should have large*

*magnetoresistance, particularly because the point contact is a completely different phase (superconducting)"*

In our arXiv paper, we have given a discussion on the difference of the PC resistance when the magnetic field increased from 0 to 3 T at temperature higher than $T_c$. Thus, there is no contribution to magnetoresistance from the superconducting phase.

Furthermore, if the magnetoresistance were due to the Sharvin term, then the Maxwell term would be even smaller from the estimation as shown in our arXiv paper, as a result, the critical current effect is even less probable. For both the PC states in Fig. 2 and 3 (arXiv: 1607.00513), the $R_m/R_{sh}$ is much smaller than 1%, so the $R_m$ is so small that it could not give the significant conductance dips in our PCS. Considering the topological properties of TaAs single crystal and the three typical features in our PCS, p-wave like superconductivity is more favorable in our case, as detailed in our paper (arXiv: 1607.00513).

*"Here we show some representative spectra where features similar to that obtained on TaAs by Wang et al. can be seen in point contacts with elemental superconductors like Nb and Pb."*

We want to emphasize the differences between the PCS in their comment and ours:

First, For most of PCS (Fig. 1a,c,d; Fig. 2a,c,d) in the comment by Gayen *et al.*, there are pronounced zero bias conductance enhancement much larger than twice of the conductance at the normal state. This over-enhancement of zero bias conductance is due to the thermal effect. We agree with that the PC of these states are in the thermal regime, because the inverted V-shape conductance feature around zero bias is a kind of typical PCS in thermal regime for conventional superconductor. Such features of the inverted V-shape and the conductance enhancement above twice of its normal state value are clearly absent in our PCS.

Moreover, the conductance dips in Figs. 1 and 2 in their comment locate at various bias positions for different PC states on the same kind of superconductors (Pb or Nb), which also indicate that the conductance dips are not intrinsic properties of the sample and is induced by the critical current effect in thermal regime. This is different with our PC results too. For our two PC states with normal state resistance 18.8 and 4.2 Ω, similar shapes and bias positions of conductance dips are shown in Figs. 2 and 3 in our arXiv paper[1], which suggests the conductance dips are intrinsic, and should be related to the superconducting order parameter.

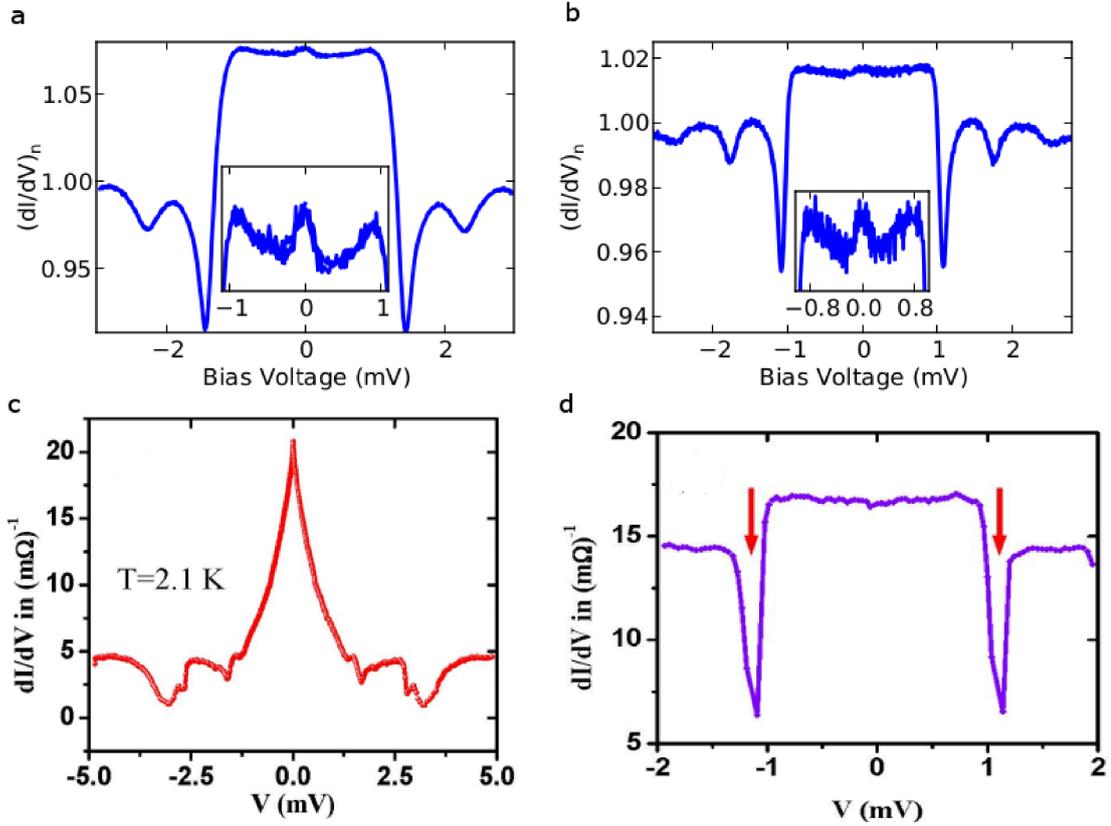

Figure 2. The clear comparison between our measurements in TaAs (a-b) and those using conventional superconductors in the comment by Gayen *et al.* (c-d). The essential difference can be found. a-b. The PCS of 18.8 and 4.2 Ω at 0.5 K in our paper (arXiv: 1607.00513), three features can be seen: zero bias conductance peak, conductance plateau with double conductance peaks at finite bias, double conductance dips. Inset: zoom-in of ZBCP and double conductance peaks at finite bias. c. The Fig.1a in the comment by Gayen *et al.*. The inverted V-shape conductance feature on Nb indicating the PC in thermal regime cannot be observed in our measurements in TaAs. d. The Fig. 2f in the comment by Gayen *et al.*. There is no zero bias conductance peak.

Finally, we point out that no PCS in their comment shows the three features simultaneously: the zero bias conductance peak, conductance plateau with double peaks, and the conductance dips (Fig. 2a, b). In our work, these three features can be consistently interpreted by a simplified model---a novel mirror-symmetry protected topological superconductor induced in TaAs, as shown by the theoretical curves in Fig. 4e in our arXiv paper. We also compare some typical PCS in their comment with ours in Fig. 2 above. The differences are obvious. Based on all the difference mentioned above, we think they have made an incorrect claim that their results in conventional superconductors are 'similar' to ours.

In conclusion, the comment by Gayen *el al.* [2] on the tip-induced superconductivity on TaAs is incorrect because: 1) the superconducting transition with temperature dependence of PC resistance is necessary for low barrier PC in ballistic regime. 2) the PCS measured and presented in their comment are qualitatively different from ours. 3) the $R_m$ is too small to induce significant conductance dips in the PCS for our PC on TaAs. Critical current effect cannot explain the PCS observed in our experiment. Instead, the p-wave like topological superconductivity proposed in our work[1] is a more plausible mechanism to understand the coexistence of zero bias peak, double conductance peaks, and double conductance dips in the PCS. In fact, the discovery alone of hard tip induced superconductivity on Weyl semimetal TaAs is extremely important since the Weyl semimetal is a natural candidate to realize a topological superconductor if superconductivity can be induced.

We thank Professor Lu, X. at Zhejiang University for the fruitful discussions.